\documentclass[conference]{IEEEtran}
\IEEEoverridecommandlockouts
\usepackage{cite}
\usepackage{amsmath,amssymb,amsfonts}
\usepackage{algorithm,algorithmic}
\usepackage{graphicx,textcomp,xcolor}
\usepackage{balance}
\usepackage{hyperref}

\usepackage{comment}

\def\BibTeX{{\rm B\kern-.05em{\sc i\kern-.025em b}\kern-.08em
    T\kern-.1667em\lower.7ex\hbox{E}\kern-.125emX}}
\begin{document}

%%%%%%%%%%%%%%%%%%% COMMENT
\iffalse
 ...
 ...
\fi
\newcommand{\POI}{{\textsc{Poi}}}
\newcommand{\POIs}{{\textsc{Poi}s}}
\newcommand{\LSTM}{{\textsc{Lstm}}}
\newcommand{\BERT}{{\textsc{Bert}}}
\newcommand{\PoiBert}{\textsc{PoiBert}}
\newcommand{\PoiLstm}{\textsc{PoiLstm}}
\newcommand{\MLM}{{\textsc{Mlm}}}
\newcommand{\NSP}{{\textsc{Nsp}}}
\newcommand{\NLP}{{\textsc{Nlp}}}

\newcommand{\spmf}{\textsc{Spmf}}

\newcommand{\Buda}{{Budapest}}
\newcommand{\Delh}{{Delhi}}
\newcommand{\Edin}{{Edinburgh}}
\newcommand{\Glas}{{Glasgow}}
\newcommand{\Osak}{{Osaka}}
\newcommand{\Pert}{{Perth}}
\newcommand{\Toro}{{Toronto}}

\newcommand{\pair}[2]{\shortstack{{#1}\\{#2}}}
\newcommand{\triplet}[3]{\shortstack{{#1}\\{#2}\\{#3}}}

\IEEEoverridecommandlockouts
\IEEEpubid{\makebox[\columnwidth]{978-1-6654-8045-1/22/\$31.00~\copyright~2022 IEEE \hfill} \hspace{\columnsep}\makebox[\columnwidth]{ }}
%%% COPYRIGHT 978-1-6654-8045-1/22/$31.00 ©2022 IEEE

\author{
%%%
%\author{\IEEEauthorblockN{ Ngai Lam Ho}
%\IEEEauthorblockA{\textit{Information Systems Technology and %Design } \\
%\textit{Singapore University of Technology and Design}\\
%\texttt{ngailam\_ho@mymail.sutd.sutd.edu.sg}}
%%% https://orcid.org/0000-0003-4768-2208
%  \and
%\IEEEauthorblockN{Kwan Hui Lim}
%\IEEEauthorblockA{\textit{Information Systems Technology and %Design }   \\
%\textit{Singapore University of Technology and Design}\\
%\texttt{kwanhui\_lim@sutd.sutd.edu.sg}}
%}
%%% https://orcid.org/0000-0002-4569-0901
%%%
    \IEEEauthorblockN{
       Ngai Lam Ho and
       Kwan Hui Lim}
    \IEEEauthorblockA{%\IEEEauthorrefmark{1}
    Information Systems Technology and Design Pillar \\
    Singapore University of Technology and Design \\
    Email:
     \href{mailto:ngailam\_ho@mymail.sutd.edu.sg}{ngailam\_ho@mymail.sutd.edu.sg},
     \href{mailto:kwanhui\_lim@sutd.edu.sg}{kwanhui\_lim@sutd.edu.sg}
    }
}

\title{ 
  %% A Transformer-based Model on Tour Recommendation Problem using~\POI-embedding
  \PoiBert: A Transformer-based Model for \\
  the Tour Recommendation Problem
  %\thanks{Identify applicable funding agency here. If none, delete this.}
}

\IEEEpubidadjcol

%%% COPYRIGHT 978-1-6654-8045-1/22/\$31.00~\copyright~2022 IEEE
\IEEEoverridecommandlockouts
\IEEEpubid{\makebox[\columnwidth]{978-1-6654-8045-1/22/\$31.00~\copyright2022 IEEE \hfill}
\hspace{\columnsep}\makebox[\columnwidth]{ }}

\maketitle

\begin{abstract}
Tour itinerary planning and recommendation are challenging problems for tourists visiting unfamiliar cities.
Many tour recommendation algorithms only consider factors such as the location and popularity of Points of Interest (POIs) but their solutions may not align well with the user’s own preferences and other location constraints. 
Additionally, these solutions do not take into consideration of the users' preference based on their past POIs selection.
In this paper, we propose~\PoiBert, an algorithm for recommending personalized itineraries using the \BERT~language model on POIs. 
\PoiBert~builds upon the highly successful~\BERT~language model with the novel adaptation of a language model to our itinerary recommendation task, alongside an iterative approach to generate consecutive POIs.

Our recommendation algorithm is able to generate a sequence of POIs that optimizes time and users' preference in POI categories based on past trajectories from similar tourists.
Our tour recommendation algorithm is modeled by adapting the itinerary recommendation problem to the sentence completion problem in natural language processing (NLP). We also innovate an iterative algorithm to generate travel itineraries that satisfies the time constraints which is most likely from past trajectories.
Using a Flickr dataset of seven cities, experimental results show that our algorithm out-performs many sequence prediction algorithms based on measures in recall, precision and \textsl{$F_1$}-scores.
\end{abstract}


%%%%%%%%%%%%%%%%%%%%%%%%%%%%%%%%%%%%%%%%%%%%%%%%%
\section{Introduction}
  Tour recommendation and planning are challenging problems faced by many tourists, due to the constraints in time and locality; additionally they may not be familiar with the city or country~\cite{brilhante-ipm15,chen-cikm16,gionis-wsdm14}.
  Most visitors usually follow guide books/websites to plan their daily itineraries or use recommendation systems that suggest places-of-interest (POIs) based on popularity~\cite{lim2019tour}. However, these are not optimized in terms of time feasibility, localities and users’ preferences~\cite{he2017category,lim2019tour}.

  In recent years, the Transformer model has become the state-of-the-art solution for many NLP tasks. Compared to other architectures, such as \emph{Recurrent Neural Network}~(\textsc{Rnn}) and~\textsc{Lstm}, a Transformer-based model processes the \emph{entire input data} all at once.
  Additionally the \emph{attention mechanism} provides the \emph{context} for any position in the input sequence, allowing more \emph{parallelism} with good performance quality; hence less time is required for training and optimization are needed~\cite{attention_2017}.
  
\begin{figure*}[h]
    \centering
    \includegraphics[width=17.0cm]{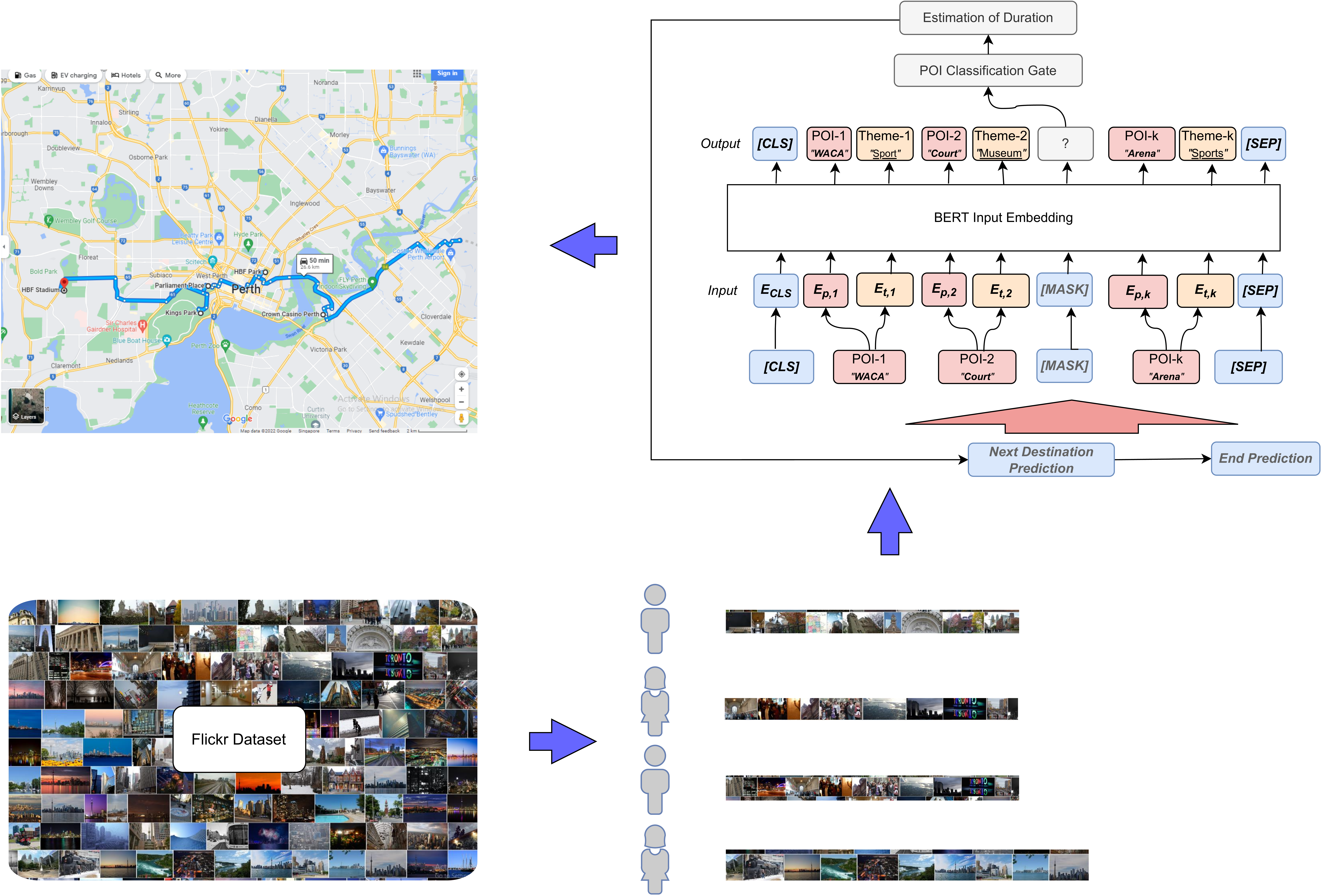}
    \caption{
      Overall system diagram of~\PoiBert~using geo-tagged photos.
      Step~(I) Given a city of interest, a set of photos with known user-IDs, timestamps and geo-tags are collected from Flickr database. Identify the {{\POI}-ID} of each photo by its geo-tag information and metadata~\cite{geotag_2015}.
      Step~(II) Sort the photos by timestamps and user-IDs to reconstruct users' trajectories to form sequences of $({\POI}-IDs , timestamps)$ tuples.
      Step~(III) Training of~\PoiBert~model using trajectories in Step-II and Algorithms~\ref{alg:mlm_data_generation}, details in Section~\ref{fig:mlm_training_themes}.
      Step~(IV) Prediction of tour itineraries using a $(source,dest.)$~\POI~tuple. 
    }
    \label{system}
 \end{figure*}

  In this paper, we propose~\PoiBert, a Transformer-word embedding model to recommend~\POIs~as a sequence of itinerary based on historical data with consideration of the locations, and also traveling time between these~\POIs. Figure~\ref{system}~shows the overall workflow of itinerary prediction of the~\PoiBert~model.

  We compare our proposed methods with other sequence prediction algorithms and conclude that our algorithms can achieve an average of~$F_1$-scores of up to 59.2\% accuracy in our experiments. In this paper, we make the following contributions:

  \begin{itemize}
      
     \item We model our Tour Recommendation problem as a \emph{sequential recommendation problem} in reinforcement learning:
     to recommend the subsequent~\POIs~(\emph{items})~in a  user's travel schedule, given a set of trajectories in the form of $user-\POI$~tuples~(\emph{item}) of interactions(\emph{check-in records})~\cite{ijcai2019_seqrec}. 
     The solution of this problem is a reinforcement learning algorithm that is flexible in different environments~(i.e. cities.)
     \item  We propose two approaches to solving the tour recommendation problem, namely:~(1)~\PoiLstm~-~A Long Short-Term Memory framework, and, (2)~\PoiBert~- a Transformer-based mechanism. 
     These two models take users' trajectories as inputs and process as a long sequence of \emph{user-item} interaction for our~recommendation~algorithms.
     \item We use the Bootstrapping method in statistics to estimate the duration of visits with \emph{confidence intervals} using a method of \emph{random sampling}. More \emph{accurate} estimation~(with confidence intervals) of~\POI~duration also results in a realistic and compact scheduling of itineraries.
     \item We have conducted thorough experiments on our proposed solutions against state-of-art solutions in sequence prediction and classic algorithms~(\textsc{Spmf}~Data Mining Library\cite{spmf2017}.) Experimentation results show that our solution out-performs other baseline algorithms.
     \item Additionally, our proposed solution has the advantage of adapting to different scenario~(cities/datasets) without modification. In particular, we recorded a performance increase, as much as 19.9\%  in our \emph{Delhi} dataset, measured in terms of average $F_1$-scores.
  \end{itemize} 

  The remaining of this paper is organized as follows:
  In Section~\ref{section_related_work} we give a background to the Tour Recommendation and discuss the state-of-the-art to the itinerary prediction problem.
  In Section~\ref{section_formulation} we formally define the problem and  notations to our  solution. 
  In Section~\ref{section_experiments} we describe our experiment framework and other baseline algorithms we used for solution evaluation. Finally, we summarize the paper with further work of extension  in~Section~\ref{section_conclusion}.
%%%%%%%%%%%%%%%%%%%%%%%%%%%%%%%%%%%%%%%%%%%%%%%%%

\section{Related Work}
    \label{section_related_work}
    \subsection{Tour Recommendation}
        \label{RELATED_TOUR_RECOMENDATION}
    Tour planning is an essential, but tedious task for tourists visiting an unfamiliar city.
    Many visitors often get recommendation from  guide books or websites to plan their daily itineraries; this will be time-consuming and sub-optimal.
    Next~\POI~prediction~\cite{he2017category,zhao2020go} and tour planning~\cite{sohrabi2020greedy,lim2019tour} are two related problems: Next-location prediction and recommendation aim to identify the next~\POI~that is most likely to visited based on historical trajectories.

    Personalized tour recommendation has been proposed with the use of photos and their meta-information such as timestamps and GPS-locations provided by Location-based Social Networks~(LBSN). Tour itinerary can be generated based on \emph{user interests} from his/her visit history. Previous works focus on recommending \emph{popular} POIs in terms of posted photos with geo-tags~\cite{geotag_2015,de2010automatic,de2010constructing,halder2022poi}. Other works utilized geo-tagged photos posted in LBSN to determine~\POI~related information for making various types of tour recommendation~\cite{lim2018personalized,cai2018itinerary,kurashima2013travel,sun2017tour,geotagged_traj_2018,halder2022efficient}.

    Furthermore, tour itinerary recommendation has the challenges of planning a \emph{connected} itinerary of~\POIs~that appeal to the users' interest preferences, without users taking unnecessary routes~ and spending extra time/distance. At the same time, there is a need to satisfy tourists' temporal and spatial constraints such as limited time and budget. 
    
    \subsection{Sequence Prediction}
    \label{RELATED_SEQ_PREDICTION}

    Sequence prediction is a well-studied machine learning task; this task involves predicting the next symbol(s) or word based on the previously observed sequence of symbols. Sequence prediction can be applied to solve the tour recommendation problem, by treating~\POIs~as words as inputs. 
    % It imposes that the order in the data must be preserved when training models and making predictions

    Sequence Prediction is widely used in the areas of time-series forecasting and product recommendation. It is different from other prediction algorithms; the order of sequence is important to get an accurate result, examples include predicting stock prices\cite{lstm_stock_2017}.
    Existing solutions to Sequence Prediction include word-embedding by considering~\POI-to-\POI~similarity using techniques such as Word2Vec, GloVe and FastText~\cite{Ayyadevara2018,bojanowski2017enriching,pennington2014glove,word2vec_rec_2020}.
    Many recommendation systems for planning tours consider broad~\POI~categories but some of their recommendations do not align well with travelers' preferences together with locational constraints in their trips. 
    Some recommendation system dynamically propose routes by taking into consideration all possible solutions generated by different agents system~\cite{Liu-ECMLPKDD20}.

    Personalized recommendation of tour planning is proposed by using the technique of {\POI}-embedding methods providing a finer representation of~\POI~categories~\cite{Lim2016PersTourAP}. 
    % In 2022, clustering-based modeling algorithm is introduced construct network topology  based on different types of n-grams on co-occurrence and word embedding distances to achieve high accuracy~\cite{topicmodel2022}.
    Recently, advances in Machine Learning (ML) and Artificial Intelligence (AI) algorithms allow for more advanced representation of sequential data, particularly in the area of Natural Language Processing.

    \subsection{\LSTM~models}
    \label{RELATED_LSTM}
    First proposed in 1994, the \emph{Long Short-Term Memory} is an \textsc{Rnn} with long-term dependencies in the input sequences\cite{lstm_1997}. 
    A \LSTM~network consists of memory blocks, or \emph{cells}, which are capable of storing information known as \emph{states}.
    During the training phase of \LSTM, two \emph{states} are transferred to~(or from, respectively) the next~(prior, respectively) cell, known as the \emph{cell state} and the {hidden state}.
    The memory blocks of \LSTM~are used as the \emph{memory} and the flow~and termination of information is done through three \emph{gates}:

    \begin{enumerate}
        \item \emph{Input Gate: }
            it is responsible for the addition of information to the \emph{cell state}. The gate applies the \emph{sigmoid} function to the input state determine information to be added to the cell state.

        \item \emph{Forge Gate:}
            this gate determines a subset of information to be \emph{removed} from the cell state; information that is less importance is removed by applying a \emph{filter} function\cite{lstm_forgetgate}. 

        \item \emph{Output Gate:}
            The Output~gate organizes the information for the output to other \LSTM~cells. The basic implementation of~\LSTM~applies the $tanh$~function cell state and the $sigmoid$ for filtering of information. The output of this gate is subsequently fed as the \emph{input gate} of the next state. 
    \end{enumerate}

    The input layer of the \LSTM~network takes in as input a vector of a \emph{fixed length} and output a vector of fixed length. In an extension of~\LSTM, the Encoder-Decoder~\LSTM~has two more additional components then the basic \LSTM~network: the \emph{encoder} and \emph{decoder}.
    The \emph{encoder} of the model extracts a \emph{fixed-length vector representation} from a variable-length input sentence.
    Experiments suggest that the encoder-decoder~\LSTM~model performs well on~\emph{short sentences} without \emph{unknown} words.
    The performance of the {\LSTM}~method was shown to \emph{degrade} as with input text size~\cite{lstm_properties_2014, seq2seq2014}.

    \subsection{Transformer models}
        Transformer is a learning model designed to process sequential input data. It adopts the mechanism of \emph{self-attention} having use primarily in~{\NLP}~and Computer Vision\cite{attention_2017}.
        Bidirectional Encoder Representations from Transformers (\BERT)  is a transformer-based machine learning technique, developed by Google\cite{bert } for \emph{language translation}. \BERT~models have become the state-of-art baseline in~{\NLP}~experiments.
        \BERT~is trained using 1) Masked-Language Modeling (\MLM), and, 2) Next Sentence Prediction~({\NSP}) with more application other than its original language tasks. Moreover,~\BERT~is shown to achieve high accuracy in \emph{Classification} tasks such as sentiment analysis~\cite{HAN2021225}.

\section{Problem Formulation and Algorithms}
    \label{section_formulation}
    In this section, we start with the definition of tour recommendation problem and a list of notations used in Table~{\ref{tbl:notations}}. 
    ~
    Given a set of travelers, $S_h$, visiting a city with $|P|$ points-of-interest, we denote a traveler, $u \in U$, in a sequence of $(poi,time)$~tuples,~$S_h = [ (p_1,t_1),(p_2,t_2)...$ $(p_k,t_k)]$, where $k$ is the number of check-in or photos posted to LBSN, for all~$p_i \in~\POIs$ and ${t_i}$ as the timestamps of the photos taken.
    Given also, a starting~\POI-${s_0} \in {\POIs}$ together with all the photos taken at~${s_0}$, the problem in this paper is to  recommend a sequence of~\POIs~in which travelers are \emph{likely} to visit based on the past trajectories from a dataset collected, using the~\emph{Transformer} model.

    We first propose ``\PoiLstm'', an {\LSTM}~model that encodes users' trajectories with consideration of the travelers' locations and distances traveled to recommend a tour itinerary with estimated duration.
    We also propose ``\PoiBert'', an algorithm for prediction of itinerary based on the \MLM~algorithm in~\BERT, discussed in~Section~\ref{bert_algo}.

        %%%%%%%%%%%%%%%%%%%%%%%%%%%%%%%%%%%%
        \begin{table}
            \centering
            \caption{Notations used in this paper}
            \scalebox{1.1}{
                \begin{tabular}{
            ||c|l||}
             \hline
                Notation & Description \\
             \hline
                $T$ & Time budget of recommended trajectory\\
             \hline
                $u_i$ & Identifier of the user ID~$i$ \\
             \hline
                $c_i$ & Category~label~(or Theme) of POI-$p_i$, e.g. Museum, \\
                ~  & Park, Sports,... \\
             \hline
                $p_i$ & Identifier of the POI ID~$i$ \\
             \hline
                $p^u_j$ & Identifier of the POI ID~$j$ in Step-$j$ of $u$'s trajectory \\
             \hline
                $v^{u}_{i}$ & Activity of user-$u$ in step-${i}$ in her/his trajectory  \\
             \hline
                 $tryj_u$ & sequence of check-ins from user-$u$ as a trajectory, \\
                ~        & i.e.  $\{ v^{u}_{1}..v^{u}_{k} \}$ \\
             \hline
                $\oplus$ & Concatenation operation \\
             \hline
                ~ & Sample distributions \\
                $X$  & $X = \{x_1,x_2,..\}$ \\
             \hline
                ~ & Empirical distributions \\
                $F^{\*}$ & $F^{\*} = \{x^{*}_1,x^{*}_2,..\}$ \\
             \hline
                $B$ & Number of sampling iterations in Bootstrapping \\
             \hline
                $\alpha$ & Significance level in Bootstrapping \\
             \hline
                \end{tabular}
            }
            \label{tbl:notations}
        \end{table}

    \subsection {\PoiLstm - Itinerary Prediction Algorithm using \LSTM}
    \label{lstm_algo}
    We model the itinerary prediction problem as a prediction in an Encoder-Decoder~\LSTM.
    Each input vector to the \PoiLstm~network represents a vector representing of user's visit~from a~\POI~transiting to the next~\POI~(with embedded details, such as \emph{time} and \emph{distance} traveled).
    During the training phase of \PoiLstm, each~\POI~in a trajectory is passed to the input layer of the \LSTM~network as an encoded vector, one at a time using the encoder function. This process is repeated for all~\POIs~in all trajectories in the training dataset, discussed in Figure~\ref{lstm_encode_function}. 
    When the \LSTM~network is trained sufficiently for a number of steps~(\emph{epochs}), the output of the \LSTM~network is a prediction of the next~\POI~(as one-hot embedding) and its estimated duration (in hours) in floating point format.)
    \POI~itinerary can be predicted by repeatedly decoding the output of \PoiLstm, by passing in the previous output information of trajectory iteratively as an \emph{encoded vector}. 

    The function $time(i,j)$ returns the time spent from $v_i$ to $v_j$ and $dist(i,j)$ returns the distance the user($u$) traveled from step-$i$ through step-$j$. Additionally, $ p^{u}_{t_{k-2}} $ , $p^{u}_{t_{k-1}}  $ and $p^{u}_{t_{k}}$ are represented as \emph{onehot embedding}~\cite{onehot}.
    
    \begin{algorithm}
      \caption{Prediction  model in \PoiLstm}
      \label{alg:PoiLstm}
      \label{lstm_encode_function}
      \begin{algorithmic}[1]
        \REQUIRE  $ v^{u}, TimeLimit$ : time budget\\
        \STATE  \textbf{Set} Activation Function: \textbf{Softmax} \\
        \STATE  \textbf{Set} Optimizer: \textbf{RMSprop} \\
        \STATE  \textbf{Let} $i=1$, $T=0$, $seq=\{\}$ \\
        \STATE  \textbf{SubFunction:} $LSTM\_encode\_seq( v^u_{t_k} ) =$ \\
        \STATE  $ ~~~~~ { time(k-1,k) } \oplus { time(1,k) } ~\oplus $ \\
        \STATE $ ~~~~~{ dist(k-1,k) }~ \oplus~dist(1,k)~\oplus $
        \STATE $~~~~~   \
              p^{u}_{t_{k-2}} \oplus \
              p^{u}_{t_{k-1}} \oplus \
              p^{u}_{t_{k}}$ 
            %\end{dmath}

        \REPEAT
            \STATE ~~$x^{(t)}_i \gets LSTM\_encode\_seq( v_{p^{u}=p_i} ) $\footnote{refer to \ref{lstm_encode_function}.} \\
            \STATE ~~compute $a^{(t)}$ and $h^{(t)}$\\
            \STATE ~~compute $o^{(t)}$ \\
            \STATE ~~\textbf{Let} $(p_{i},t_{i}) \gets decode( o^{(t)} )$
            %% \STATE Visit \POI-(p_{i})   $
            \STATE ~~$seq \gets seq \oplus p_{i}$
            \STATE ~~$T \gets T+t_{i}$
            \STATE ~~$i \gets i+1$
        \UNTIL{  $ T  \ge TimeLimit$ } 
        \RETURN $seq$ \\
            %\begin{dmath}
      \end{algorithmic}
    \end{algorithm}

    \subsection{\PoiBert~-~a~\BERT~model for~\POI~Itinerary Prediction} \label{bert_algo}
        Generally, a \BERT~model uses a \emph{self-attention} mechanism that is able to  learn the general trend of a language; the trained model can then be used for downstream tasks, such as \emph{language translation} and \emph{question answering}~\cite{bert_question_answering}. When used in practice, a pre-trained \BERT~model can significantly improve the results in prediction based on a number of benchmarks. 
        To perform an itinerary prediction in our \PoiBert~model, we pass in a set of \emph{sentences} consisting of~\POIs~and relevant information to predict the next~\POI~which is most likely to occur using the \MLM~prediction model.
    
    \begin{paragraph}{Training of \PoiBert~Model}
      We propose a novel~\PoiBert~Model in the space of~\POIs~and users' itineraries. 
      The original implementation of \BERT~train~\MLM~ by masking 15\% of words.
      The \PoiBert~prediction algorithm is to predict the \emph{masked}~\POI~(word), based on the context provided by other \emph{words}~(representing~\POIs~or~\POI~categories)  \emph{without~masks}.
      We use Algorithm~\ref{alg:mlm_data_generation} to translate users' trajectories into sentences of~\POIs(\emph{words}) which are subsequently trained by the~\PoiBert~model for the itinerary prediction task. 
     
      Figure~\ref{alg:mlm_data_generation} outlines a function to transform users' trajectories to sentences of words representing~\POIs~or categories of~\POIs~for~\PoiBert~training.
      The time complexity of the function is $O(N K^2)$, where~$N$~is the total number of~\POIs~in the dataset and~$K$ represents the maximum number of~\POIs~in any trajectory.
    
        %%% FIGURE 2
        \begin{figure*}[t]
          %%% FROM BERT_with_themes.drawio.drawio
          \label{fig:mlm_training_themes}
          \centering
          %%\includegraphics[width=0.99\linewidth]{BERT_with_themes.drawio}
          % BERTsystem
          \includegraphics[trim=48cm 31cm 0cm 0cm,clip,width=15.5cm]{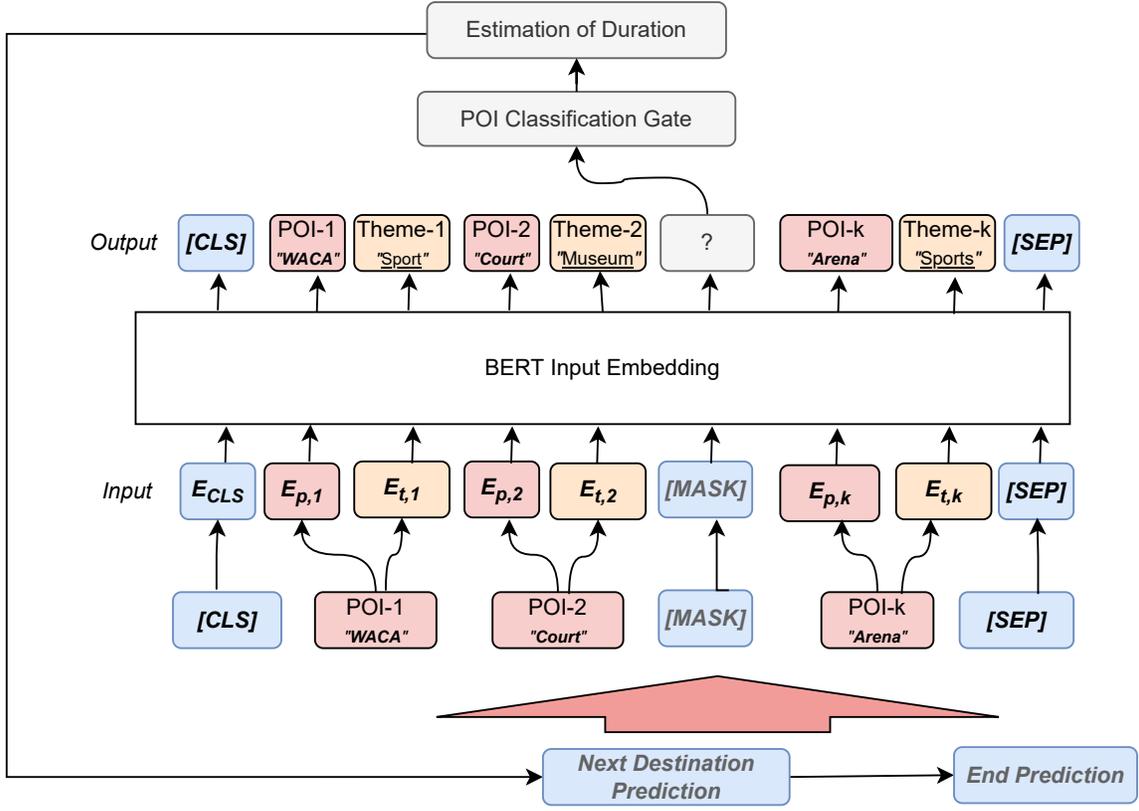}
          
          %% BERT_with_themes.drawio
          \caption{Itinerary prediction algorithm of \PoiBert~model:  In each iteration of the system, a new destination~(i.e.~\POI) is predicted by solving the~\MLM~prediction task; the predicted~\POI~is then inserted to the itinerary. 
          The prediction loop stops when all~\POIs~are visited, or when the time constraint is satisfied.}
        \end{figure*}
        
        \begin{algorithm}
          \caption{Training Data Generation~for~\PoiBert}
          \label{alg:mlm_data_generation}
          \begin{algorithmic}[1]
            \REQUIRE  $ tryj_u , \forall u \in Users $ \\
            \FORALL{ $u \in users$ }
                %\STATE line for
                \FORALL{ $tryj\_seq \in tryj_u$}
                    \STATE \textbf{Let} $n \leftarrow | tryj\_seq | $ \\
                    \STATE \textbf{Let} $\{p_1..p_n\} \leftarrow  poi\_id(tryj\_seq)$ \\
                    \STATE \textbf{Let} $\{c_1..c_n\} \leftarrow  theme(tryj\_seq)$ \\
                    \STATE  // where the functions $poi\_id(...)$ and $theme(...)$ \\
                    \STATE  // ~return $POI\_id$~(and $theme$, resp.) projections.   \\
                    \STATE \textbf{Output:} $\forall 1 \le i < j \le n$,\\
                    \STATE ~~~~~~~~~~~``$ \{  c_i,p_i,..,p_{j-1},c_{j-1} \} \rightarrow p_j$''
                \ENDFOR
            \ENDFOR
            %\RETURN $seq$
          \end{algorithmic}
        \end{algorithm}

    \end{paragraph}

    \begin{paragraph}{Itinerary Prediction}
        Given an initial POI, $p_1$, and the ending POI, $p_k$ from traveler's specification, we propose an algorithm to predict a sequence of~\POIs~which travelers are most \emph{likely} to visit as ordered list, based of historical trajectories recorded in the dataset.
        The \PoiBert~algorithm is inspired by the~\MLM~training process of \BERT, where the prediction algorithm identifies the \emph{masked} words based on the context of a sentence.
        As outlined in Algorithm~\ref{alg:PoiBert}, the algorithm  \emph{searches} for the next relevant~\POI~between the initial~\POI~and destination~\POI, and insert it to the predicted itinerary.
    
        \begin{algorithm}
          \caption{Itinerary Prediction Algorithm in \PoiBert}
          \label{alg:PoiBert}
          \begin{algorithmic}[1]
            \REQUIRE  $p_{1},p_{k}$: starting/ending~\POIs \\
               ~~~~~~~$TimeLimit$:  time budget of itinerary
            \STATE  \textbf{Let} $seq \gets \{ p_{1},p_{k} \}$ \\
            \REPEAT
                \STATE \textbf{forall}~ {$j \in \{2..|seq|-1\}$} \\
                {
                \STATE ~~~\textbf{Let} $query_j \gets \{ p_{1},c_{1},p_{j-1},c_{j-1},            \texttt{{[MASK]}},$\\
                ~~~~~~~~~~~~~~~~~~~~~~~$p_{j},c_{j},...,p_{k},c_{k} \}$ \\
                }
                \STATE $seq \gets \textbf{ArgMax}_{j\in\{2..|seq|-1\}}
                        ( \textbf{\textit{Unmask}}(query_j)) $ 
            \UNTIL{  $ \displaystyle \sum_{poi \in seq}{duration( poi )} \ge TimeLimit$ } 
             \RETURN $seq$
          \end{algorithmic}
        \end{algorithm}

    \end{paragraph}
    
    \subsection{Estimation of duration of visits}
    Getting a realistic estimate of duration of visits to our predicted~\POIs~are crucial in our solution. Any over-estimation (or  under-estimation) of duration to the predicted~\POIs~will affect the time-sensitive  trajectories output from the algorithm, hence affecting the recall- and precision-scores. In this section, we estimate the duration of visits using a statistical method:~\emph{bootstrapping} by calculating the \emph{confidence-interval} of duration in the trajectories~\cite{2010statistics}. 
    Due to the high \emph{variance} in duration of visit to the~\POIs, it is not practical to estimate the duration by merely taking the \emph{averages} of all visitors' duration to the~\POIs. 
    
    We note that Bootstrapping does not assume the input data to follow any statistic distribution.
    It treats the original \emph{samples} as the \emph{real population} and draws random samples from the data. Then, bootstrapping creates \emph{more} samples so one can make a better estimation of population by the procedure of \emph{sub-sampling randomly with replacement}. This method is used to estimate the duration of visits at the~\POIs~given that there are less samples visiting to some~\POIs~that are less popular. Algorithm~\ref{alg:bootstrapping} outlines the steps of getting the 90\% confidence~intervals of duration of
    visit to a \POI-$i$, $\forall i \in~\POIs$.

   \begin{algorithm}
      \caption{Estimate Duration of Visit to \POI~}
      \label{alg:bootstrapping}
      \begin{algorithmic}[1]
        \REQUIRE  $ poi\_id \in \POIs$ \\
        \REQUIRE  confidence~level~$\alpha$
        \REQUIRE  number of replicates~$B$
        \REQUIRE  $ Tryj_u , \forall u \in Users $ \\
        \STATE \textbf{SubFunc.} $getSamples(poi\_id)$:
            \FORALL{ $u \in users$ }
            \FORALL{ $tryj\_seq \in tryj_u$} 
                \FORALL{ $p \in tryj\_seq $}
                    \STATE \textbf{Output}  activities if $p == poi\_id $ \\
                \ENDFOR
            \ENDFOR
        \ENDFOR~
        \STATE ~\\  
        \STATE \textbf{Let}~$X \gets getSamples(poi\_id) $\\
        \STATE \textbf{Sample} $x^*_1, x^*_2,..x^*_n$ \text{with replacement from sample}~$X$.\\
              \textbf{Repeat}~$B$ iterations\footnote{$B$ is a large number for bootstrapping to be efficient, we use $B=10000$ for all experiments}. 
        \STATE \textbf{Let}~$F^*$ be the \emph{empirical distribution}
        \STATE Calculate $(100-\alpha)$\% \emph{confidence intervals} for~$F^*$ 
        \end{algorithmic}
    \end{algorithm}

\section{Experiments and Results}
    \label{section_experiments}
    We use a dataset of photos uploaded to the Flickr platform, which consists of trajectories of 5,654 users from 7 different cities, tagged with meta-information, such as the date and GPS location. Using this data-set, one can construct the travel trajectories by sorting the photos by time, mapping the photos to the~\POIs~as identified by the GPS location, resulting in the users’ trajectories as a sequence of time sensitive \POI-IDs. 
                
    %%% Dataset info
    \begin{table*}[h]
        \centering
        \caption{Description of Data-sets}
          \scalebox{1.30}{
%          \resizebox{\linewidth}{
            \begin{tabular}{lccccccccc }
              \hline\hline
                City & \Buda & \Delh & \Edin & \Glas & \Osak &  \Pert & \Toro
              \\
              \hline\hline
            {No. of~\POIs} & 39 & 26 & 29 & 29 & 28 & 25 & 30 \\ % & 29 \\
              \hline
            No. of Users &  935 sample & 279 & 1454 & 601 &  450  & 159 &  1395 \\ % &  1155 \\
              \hline
            No. of Trajectories & 2361 & 489 & 5028 & 2227 & 1115 & 716 &  605 \\ %& 3193 \\
            .. for training~~~~ & 277 & 13 & 267 & 28 & 12 & 14 & 95 \\
            .. for evaluation & 70 & 4 & 67 & 7 & 3 & 4 & 74  \\
              \hline
            No. of check-ins~~~~ & 18513 & 3993 & 33944 & 11434 & 7747 & 3643 & 39419 \\% & 34515 \\
            .. for training~~~ & 7593  & 367 & 6575 & 600 & 381 & 418 & 2371 \\% & 6640 \\
            .. for evaluation & 1915 & 215 & 1921 & 223 & 40 & 68 & 540 \\% & 2536 \\
          \hline
        {Avg.~\POIs~per~trajectory} & 5.78  & 4.69 & 5.27 & 4.71 & 4.50 & 4.93 & 4.93 \\% & 5.271 \\
          \hline
        {Avg.~check-ins~per~\POI~~}  &   4.74 & 6.02 & 4.68 & 4.55 & 7.06 & 6.06 & 5.07 \\% & 5.338  \\
          \hline\hline
            \end{tabular}
          }
        \label{fig:datasets}
    \end{table*}

    %%% F_1 / RECALL/ precision scores table
    %%% Average~{$F_1$} / Recall / Precision scores of \PoiBert~prediction algorithm~(\%) }
    \begin{table*}[!th]
        \centering
        \caption{Average~{$F_1$} / Recall / Precision scores of \PoiBert~prediction algorithm~(\%) }
        \scalebox{1.32}{
%          \resizebox{\linewidth}{
            \begin{tabular}{        cccccccccccc}
            \hline\hline
          epochs & ~ & \Buda & \Delh & \Edin & \Glas &  \Osak &  \Pert & \Toro \\
            \hline\hline
            \triplet{~}{1}{~~} &
                \triplet{\it Recall}{$F_1$}{\it Precision} &
                \triplet{65.511}{\bf 49.974}{46.563}  & \triplet{71.250}{58.485}{50.119}  & \triplet{64.262}{\textbf{54.371}}{53.223}  & \triplet{78.265}{\textit{\underline{59.417}}}{52.234}  & 
                \triplet{46.667}{52.382}{61.111}  &   \triplet{77.500}{60.242}{54.924}  &   \triplet{73.427}{\textbf{55.929}}{52.559} \\
            \hline
            \triplet{~}{3}{~~} &
                \triplet{\it Recall}{$F_1$}{\it Precision} &
                \triplet{63.498}{\textit{\underline{48.533}}}{45.632}  & \triplet{87.500}{58.485}{55.357}  & \triplet{64.165}{\textbf{54.371}}{52.950}  & \triplet{81.020}{59.381}{52.531}  & 
            \triplet{55.000}{52.381}{75.556}  &   \triplet{77.500}{60.242}{60.417}  &   \triplet{73.427}{\textbf{55.929}}{50.666} \\  
            \hline
            \triplet{~}{5}{~~} &
                \triplet{\it Recall}{$F_1$}{\it Precision} &
                \triplet{60.455}{47.448}{45.238}  & \triplet{76.250}{58.333}{47.619}  & \triplet{61.965}{52.748}{52.694}  & \triplet{81.020}{\textbf{60.752}}{53.296}  & 
                \triplet{54.999}{63.420}{75.556}  &   \triplet{77.500}{\textit{\underline{61.994}}}{61.174}  &   \triplet{74.468}{52.973}{52.618} \\
            \hline
            \triplet{~}{7}{~~} &
                \triplet{\it Recall}{$F_1$}{\it Precision} &
                \triplet{63.094}{\it 48.731}{46.014}  &
                \triplet{76.250}{58.333}{47.619}  &
                \triplet{61.710}{52.229}{51.909}  &
                \triplet{70.306}{54.949}{48.044}  & 
                \triplet{55.000}{63.420}{75.556}  &
                \triplet{77.500}{60.242}{60.417}  &
                \triplet{71.790}{52.256}{52.856} \\
            \hline
            \triplet{~}{10}{~~} &
                \triplet{\it Recall}{$F_1$}{\it Precision} &
                \triplet{61.323}{47.542}{45.425}  &
                \triplet{76.250}{58.333}{47.619}  &
                \triplet{62.148}{53.397}{53.145}  &
                \triplet{76.735}{51.042}{47.086}  & 
                \triplet{61.667}{\textit{\underline{71.753}}}{86.667}  &
                \triplet{72.500}{\bf 64.286}{52.083}  &
                \triplet{64.865}{52.744}{52.825} \\
            \hline

            \triplet{~}{15}{~~} &
                \triplet{\it Recall}{$F_1$}{\it Precision} &
                \triplet{60.717}{46.884}{44.510}  &
                \triplet{76.250}{58.333}{47.619}  &
                \triplet{62.507}{\textit{\underline{53.556}}}{53.206}  &
                \triplet{66.225}{51.471}{45.899}  & 
                \triplet{53.333}{60.714}{72.222}  &
                \triplet{72.500}{55.777}{53.750}  &
                \triplet{67.782}{54.589}{54.592} \\
            \hline

            \triplet{~}{20}{~~} &
                \triplet{\it Recall}{$F_1$}{\it Precision} &
                \triplet{62.870}{48.228}{45.517}  &
                \triplet{76.250}{57.051}{46.230}  &
                \triplet{60.855}{51.865}{51.064}  &
                \triplet{78.980}{56.724}{48.566}  & 
                \triplet{70.000}{\bf 74.817}{84.127}  &
                \triplet{66.250}{56.047}{54.464}  &
                \triplet{64.320}{50.288}{49.533} \\
            \hline
            \triplet{~}{30}{~~} &
                \triplet{\it Recall}{$F_1$}{\it Precision} &
                \triplet{60.469}{47.081}{45.167}  &
                \triplet{76.250}{58.333}{47.619}  &
                \triplet{61.611}{51.806}{53.215}  &
                \triplet{73.367}{51.752}{46.315}  & 
                \triplet{53.333}{62.229}{75.556}  &
                \triplet{66.250}{59.077}{56.548}  &
                \triplet{63.273}{52.542}{52.212} \\
            \hline

            \triplet{~}{40}{~~} &
                \triplet{\it Recall}{$F_1$}{\it Precision} &
                \triplet{59.210}{45.675}{43.258}  &
                \triplet{82.500}{63.333}{51.786}  &
                \triplet{60.991}{51.494}{51.813}  &
                \triplet{69.796}{51.696}{44.490}  & 
                \triplet{53.333}{62.229}{75.556}  &
                \triplet{61.250}{52.411}{52.381}  &
                \triplet{63.442}{50.514}{51.548} \\
            \hline

            \triplet{~}{50}{~~} &
                \triplet{\it Recall}{$F_1$}{\it Precision} &
                \triplet{60.673}{46.686}{44.280}  &
                \triplet{82.500}{\textit{\underline{64.848}}}{54.167}  &
                \triplet{60.141}{51.465}{50.924}  &
                \triplet{75.408}{53.457}{47.973}  & 
                \triplet{53.333}{62.229}{75.556}  &
                \triplet{60.000}{54.708}{50.947}  &
                \triplet{63.863}{52.506}{51.301} \\
            \hline

            \triplet{~}{60}{~~} &
                \triplet{\it Recall}{$F_1$}{\it Precision} &
                \triplet{60.453}{47.186}{45.030}  &
                \triplet{88.75}{\textbf{69.848}}{57.738}  &
                \triplet{61.445}{52.240}{51.566}  &
                \triplet{66.224}{49.128}{43.900}  & 
                \triplet{53.333}{62.229}{75.556}  &
                \triplet{66.25}{54.708}{55.159}  &
                \triplet{66.182}{51.777}{51.935} \\
            \hline\hline
            &
            \end{tabular}
            }
%        }
        \label{table:f1scores}
    \end{table*}

    \subsection{Datasets}
    We use the Flickr datasets prepared for our evaluation of  algorithms~\cite{Lim2016PersTourAP}.
    In total, there are close to 120K photos, or check-in records, from 4701 users in seven popular cities.
    Table~\ref{fig:datasets} describes more details about each dataset and information about the trajectories of these cities.

    \begin{paragraph}{Training and Test Set}
    Our data-sets are split into Training and Testing data-sets. Firstly, we organize photos by the Trajectory-IDs, then these trajectories are sorted according to their \emph{last~check-in times} (in ascending order).
    To obtain the Training dataset, the first 80\% of Trajectories~(based on their photos) are set aside as \emph{Training Data}. The remaining data is used as the \emph{Testing Data}. This segregation of Training and Test data avoids the problem of having a trajectory covering over both Training and Testing Data sets.
    \end{paragraph}

    %%% ALGORITHMS table ALL ALGOITHMS
    \begin{table*}[!th]
        \centering
        \caption{
           Average $F_1$ scores for different Sequence Prediction Algorithms~(\%) }
        \scalebox{1.3}{
        %% |p{0.10\linewidth}||c|c|c|c|c|c|c|c|c|} 
        \begin{tabular}{
        p{1.8cm}ccccccccc}    \hline
         Algorithm  & \Buda & \Delh &  \Edin & \Glas & \Osak & \Pert & \Toro \\
        \hline
        \textsc{Cpt} & 45.331 & 58.238 & 44.732 & 51.234 & 45.238 & 58.569 & 46.816  \\
        % \hline
        \textsc{Cpt+} & 43.472 & 42.511 & 44.543 & 48.701 & 37.719 & 58.570 & 37.719 \\
        % \hline
        \textsc{Dg} & 44.917 & 50.260 & 44.867 & 50.579 & 43.333 & 49.936 & 43.333 \\
        % \hline
        \textsc{Lz78} & 43.447 & 49.412 & 44.105 & 45.438 & 40.00 & 51.582 & 40.00 \\
        % \hline
        PPM & 44.574 & 50.258 & 44.848 & 50.579 & 45.556 & 54.481 & 45.556 \\
        % \hline
        \textsc{Tfag} & 43.686 & 60.694 & 43.105 & 48.237 & 45.556 & 48.711 & 45.555 \\
        % \hline
        \textsc{Bwt}-SuBSeq & 37.883 & 43.333 &  39.082 &  48.322 & 42.857 & 36.320 &  33.145 \\
        % \hline
        \textsc{Seq2Seq} & 36.970 &  43.864 & \textit{\underline{52.768}} & \textit{\underline{62.132}} & 57.937 & 54.911 & \textit{\underline{52.870}} \\
        % \hline
        \textbf{\PoiLstm}*  & \textbf{53.591} & \textit{\underline{68.750}} & 41.282 & 61.147 & \textit{\underline{60.350}} & \textit{\underline{60.229}} & 50.759 \\
        % \hline
        \textbf{\PoiBert}* & \textit{\underline{49.974}} & \textbf{69.848} & \textbf{54.471} & \textbf{62.771} & \textbf{71.753} & \textbf{61.075} & \textbf{55.929} \\
          \hline
        \end{tabular}
        }
        \label{table:all_algo}
    \end{table*}

    %%% Average Number of~\POI's using \PoiBert~Predicted Model vs. Actual Trajectories
    \begin{table*}[!th]
        \centering
        \caption{
           Average Number of~\POI's using \PoiBert~Predicted Model vs. Actual Trajectories }
        \scalebox{1.4}{
            \begin{tabular}{
            p{2.3cm}ccccccccc}
                \hline
                 Epoches  & \Buda & \Delh &  \Edin & \Glas & \Osak & \Pert & \Toro \\
                \hline
                \textsl{Actual Trajectories} & \textsl{6.243} & \textsl{4.750} & \textsl{5.955} & \textsl{5.000} & \textsl{5.000} & \textsl{5.250} & \textsl{5.458} \\ %\hline
                 1 & 9.786 & 6.000 & 7.881 & 7.429 & 4.000 & 6.750 &  7.583 \\ %\hline
                 3 & 9.814 & 6.750 & 7.582 & 7.857 & 3.667 & 7.500 & 12.042 \\ %\hline
                 5 & 9.514 & 6.750 & 7.507 & 7.714 & 3.667 & 7.500 & 11.250 \\ %\hline
                 7 & 9.729 & 6.750 & 7.881 & 7.286 & 3.667 & 7.500 & 10.917 \\ %\hline
                10 & 9.671 & 6.750 & 7.571 & 7.571 & 3.667 & 7.500 &  7.458 \\ %\hline
                15 & 9.871 & 6.750 & 7.806 & 7.000 & 4.000 & 7.000 &  7.583 \\ %\hline
                20 & 9.914 & 7.000 & 7.791 & 7.857 & 4.333 & 6.500 &  8.042 \\ %\hline
                30 & 9.757 & 6.750 & 7.672 & 6.857 & 3.667 & 5.750 &  7.250 \\ %\hline
                40 & 9.771 & 6.750 & 7.836 & 7.429 & 3.667 & 6.250 &  7.500 \\ %\hline
                50 & 9.871 & 6.500 & 7.821 & 8.000 & 3.667 & 6.250 &  7.708 \\ %\hline
                60 & 9.600 & 4.333 & 7.940 & 6.857 & 3.667 & 5.500 &  7.875 \\ \hline
            \end{tabular}
        } %%% scalebox
        \label{table:average_pois}
    \end{table*}

    \subsection{Performance of Algorithms}
        \label{accuracy}
        Experiments were conducted for each city in the dataset. We regard all users' trajectories~(with at least 3~\POIs) in the training set as sequences of~POI~(\emph{corpus}). 
        To compare the performance of our models, we trained different sequence prediction~models using different hyper-parameters. 
        We then used the Test set to evaluate the accuracy of the trained models:
        for each of the trajectory in the testing set (known as \emph{history-list}), we treat the \emph{first} (and \emph{last}, respective)~\POI~as the \emph{source} (and \emph{destination}, respectively)~\POI~and try to  predict the \emph{intermediate}~\POIs~of the trajectory, given in a time~boxed event of \emph{history-list}.
        We evaluated the effectiveness of \PoiBert~and \PoiLstm~prediction algorithms in terms of {$F_1$},
        precision~($T_{p}$)~and recall~($T_{r}$) scores of the predicted~{\POIs}~against the actual trajectories, as below:
        
        \noindent  Let $S_p$ be the predicted sequence of~\POIs~from the algorithm and $S_h$ be the actual sequence from the trajectories, we evaluate our algorithms based on: 
        
        \begin{itemize}
            \item $T_{r}(S_h,S_p)$ = $ \frac{|S_h \cap S_p|}  {|S_p|}$
            \item $T_{p}(S_h,S_p) = \frac{|S_h \cap S_p|}{|S_h|}$
            \item  $F_1\_score(S_h,S_p) =  \frac{2  T_{r}(\bullet)  T_{p}(\bullet)}
                                  {T_{r}(\bullet) + T_{p}(\bullet)}$
        \end{itemize}

    \subsection{Baseline Algorithms}
        Our proposed models are compared with other sequence prediction algorithms as baseline algorithms:
        \begin{itemize}
            \item \textsc{Spmf} algorithms - this package consists of data mining algorithms including: \emph{CPT}~\cite{CPT2013}, \emph{CPT+}~\cite{CPTplus2015}, \emph{TDAG}~\cite{DG1996}, \emph{First-order and All-k-Order } 
            \emph{Markov Chains}\cite{PPM1984,AKOM1999}. Our experiments predict an itinerary by \emph{repeatedly} asking for the next \emph{token}~(represented as the next~\POI~to visit) when time limit is not exhausted.
            \item \textsc{SuBSeq} : the algorithm uses a \emph{Succinct Wavelet Tree} structure to maintain a list of training sequences for sequence~prediction~\cite{succinctBWT_2019}.
            \item \textsc{Seq2Seq} : this model adopts a multilayered {\LSTM} to map the input sequence to a vector with a fixed size or dimension~\cite{seq2seq2014}. The prediction is facilitated by another deep {\LSTM} to decode the target sequence. The default prediction model of \textsc{Seq2Seq} is to output a \emph{sentence} of words which may consist of duplicated words. We modified the prediction model to return a series of \emph{unique}~POIs~instead.
        \end{itemize}
        
        Some baseline algorithms only predict one~\emph{token} or~\POI, we \emph{iteratively} predict more tokens until the time limit of the itinerary is reached. For the propose of algorithms evaluation, all experimentation of baseline algorithms are conducted in the same setting as in Section~\ref{accuracy}, sharing the same training and testing data.
        %%%%%%%%%%%%%%%%%%%%%%%%%%%%%%%%%%%%%%%%%%%%%%%%%
        
    \subsection{Experimental Results}
        We evaluated the effectiveness of our proposed algorithms on different cities.
        We constructed the travel histories by chronologically sorting the photos, which resulted in the users' trajectories. These trajectories are then regarded as \emph{sentences} for inputs to our proposed training models with different hyper-parameters.
        Results are summarized by comparing the accuracy of the predicted itineraries~(i.e. Recall / Precision  / {$F_1$} scores,) as shown in Table~{\ref{table:f1scores}}. 
        
        In Table~\ref{table:all_algo}, we also compare the performance~of \PoiBert~and~{\PoiBert} against  8 baseline algorithms. 
        Overall, experimental results show that our ~\PoiLstm~ and ~\PoiBert~ itinerary prediction algorithms achieve significant accuracy in itinerary prediction tasks; the proposed~\PoiBert~prediction algorithm is scale-able and adaptable in parallel environment. 
        Our experiments also show that the {\PoiBert}-prediction algorithm achieves {$F_1$~scores} of at {\textit{least}} 47\% accuracy across all cities and different parameter settings.
        In particular, we recorded an average of $74.8\%$ in our \emph{\Osak} dataset; experiments in~\emph{\Delh} also show an \emph{increase} of 19.99\%~(from 58.238\% up to 69.848\%) in~$F_1$ score.

    In Table~\ref{table:average_pois}, we compare the number of~\POIs~in users' trajectories and their  predicted itineraries by \PoiBert. \PoiBert~is able to recommend more \emph{relevant}, and \emph{compact} trajectories relative to the actual trajectories, while not compromising the quality of the recommendation model.

\section{Conclusion}
    \label{section_conclusion}
    In this paper, we study the problem of tour itinerary recommendation to identify users' preference on~\POIs~and make the appropriate recommendation of itineraries with time constraints.
    To solve this problem, we propose~\PoiBert~ that builds upon the highly successful~\BERT~ model with the novel adaptation of a language model to this itinerary recommendation task, along with an iterative approach to generating~\POIs.
    Our iterative \PoiBert~prediction algorithm can reliably uncover a user's preference in a tour by only using a pair of initial and destination~\POIs.  
    Our experiments show the effectiveness of our proposed algorithm for predicting relevant~\POIs~in terms of {\textsl{$F_1$}}-scores.
    In our experiments on 7~cities, our \PoiBert~algorithm \emph{outperforms} 8~baseline algorithms measured in~averaged~$F_1$-scores. Future works include further adaptation and more in-depth evaluation of other language models for this itinerary recommendation task and creating a \textsl{HuggingFace} interface module for~\PoiBert~\cite{huggingface2019}.

%%%%%%%%%%%%%%%%%%%%%%%%%%%%%%%%%%%%%%%%%%%%%%%%% 
\section*{Acknowledgment}
  \noindent{\small This research is funded in part by the Singapore University of Technology and Design under grant SRG-ISTD-2018-140.\\
  The computational work was partially performed on resources of the National Super-Computing Centre, Singapore.}

\bibliographystyle{IEEEtran}
\balance
\bibliography{BigD648}

\vspace{12pt}

\end{document}